\documentclass[
 reprint,
 amsmath,amssymb,
 aps,
]{revtex4-2}

\usepackage{graphicx}% Include figure files
\usepackage{dcolumn}% Align table columns on decimal point
\usepackage{bm}% bold math
\begin{document}

\preprint{}

\title{Cooperating Cracks in Two-Dimensional Crystals}% Force line breaks with \\

\author{Shizhe Feng$^{1,2}$, Xiaodong Zheng$^{3}$, Pengjie Shi$^{2}$,\\
Thuc Hue Ly$^{4}$, Jiong Zhao$^{3, \ast}$, Zhiping Xu$^{2,}$}
\email{Corresponding authors: xuzp@tsinghua.edu.cn (Z.X.), jiongzhao@polyu.edu.hk (J. Z.)}

\affiliation{$^{1}$Department of Mechanics, Sichuan University, Chengdu 610065, China \\
$^{2}$Applied Mechanics Laboratory, Department of Engineering Mechanics, Tsinghua University, Beijing 100084, China\\
$^{3}$Department of Applied Physics, The Hong Kong Polytechnic University, Kowloon 999077, Hong Kong, China\\
$^{4}$Department of Chemistry, City University of Hong Kong, Kowloon 999077, Hong Kong, China}

\date{\today}% It is always \today, today,

\begin{abstract}

The pattern development of multiple cracks in extremely anisotropic solids such as bilayer or multilayer two-dimensional (2D) crystals contains rich physics, which, however, remains largely unexplored.
We studied crack interaction across neighboring 2D layers by transmission electron microscopy and molecular dynamics simulations.
Parallel and anti-parallel (`\emph{En-Passant}') cracks attract and repel each other in bilayer 2D crystals, respectively, in stark contrast to the behaviors of co-planar cracks.
We show that the misfit between in-plane displacement fields around the crack tips results in non-uniform interlayer shear, which modifies the crack driving forces by creating an antisymmetric component of the stress intensity factor.
The cross-layer interaction between cracks directly leads to material toughening, the strength of which increases with the shear stiffness and decreases with the crack spacings.
Backed by the experimental findings and simulation results, a theory that marries the theory of linear elastic fracture mechanics and the shear-lag model is presented, which guides the unconventional approach to engineer fracture patterns and enhance material resistance to cracking.
%Keywords: 2D crystals, crack propagation, cross-layer crack interaction,  interlayer shear, shear-lag

%\begin{description}
%\end{description}
\end{abstract}

\maketitle

%\tableofcontents

Fracture is a catastrophic process that leaves cracks behind as materials fail, which could also be harnessed to fabricate structures with specific edge shapes in a positive perspective~\cite{lawn1977fracture, feng2023controlling}.
The problem of cracking has been rationalized in the theory of fracture mechanics since the seminar contribution of Griffith in 1921~\cite{griffith1921vi}.
However, the main focus of studies has been placed on single cracks nucleating and growing in a solid, although in practice, cooperative dynamics of interacting cracks are often of vital importance by defining the fracture patterns and energy dissipation processes~\cite{goehring2015desiccation,quinn2021radial}.
Model problems of two interacting cracks have been studied in the literature~\cite{broberg1999cracks}.
In-plane parallel cracks propagating in the same direction kink away from each other depending on the mixing of fracture modes
%$K_{\rm II}/K_{\rm I}$
~\cite{ramesh2008cylindrical}.
\emph{En-Passant} or anti-parallel cracks repel and then attract while approaching each other~\cite{kranz1979crack, dalbe2015repulsion, schwaab2018interacting}, which is attributed to shielding and amplification of the crack-tip stress fields~\cite{xia2000crack, ghelichi2015modeling}.

Two-dimensional (2D) crystals and their layered assemblies have drawn notable attention for promising applications in materials and devices~\cite{novoselov20162d}.
One of the unique features of these materials or structures is the high contrast between the basal-plane and interlayer mechanical properties~\cite{wang2007extreme}.
The weak van der Waals (vdW) and/or electrostatic forces transfer the mechanical load across the 2D layers such as graphene, hexagonal boron nitride (h-BN), and metal dichalcogenides (TMDs), compete with the elasticity of 2D crystals and modulate the lattice deformation~\cite{feng2021pattern}.
It is thus of both fundamental and practical interest to explore the behaviors of multiple cracks propagating in different layers, which are modulated by the interlayer interactions.
$In \ situ$ scanning electron microscopy (SEM) studies show bifurcated fracture patterns in multilayer graphene, where mechanical energy dissipation in post-fracture events of interlayer slip improves the fracture toughness~\cite{jang2017asynchronous}.
Complex crack-tip behaviors were reported to be governed by highly variable interlayer friction.
Cracks advancing across a bilayer MoS$_2$ can either propagate, be blocked, or branch due to the combined effects of lattice defects and interaction between the layers.
Cracks propagating in one layer can even break the other layer with small flaws located in their paths~\cite{jung2018interlocking}.
These experimental phenomena imply the significance of interlayer coupling in defining fracture patterns in layered materials.

Here, we carried out an $in~situ$ scanning transmission electron microscopy (STEM) study of the interaction between cracks advancing in neighboring layers of 2D crystals.
The underlying mechanisms governed by interlayer strain transfer are explored by atomistic simulations and elucidated continuum mechanics analysis, which offer an unconventional approach to enhance crack resistance and material toughness.

\begin{figure}[h]
\includegraphics[width=\linewidth]{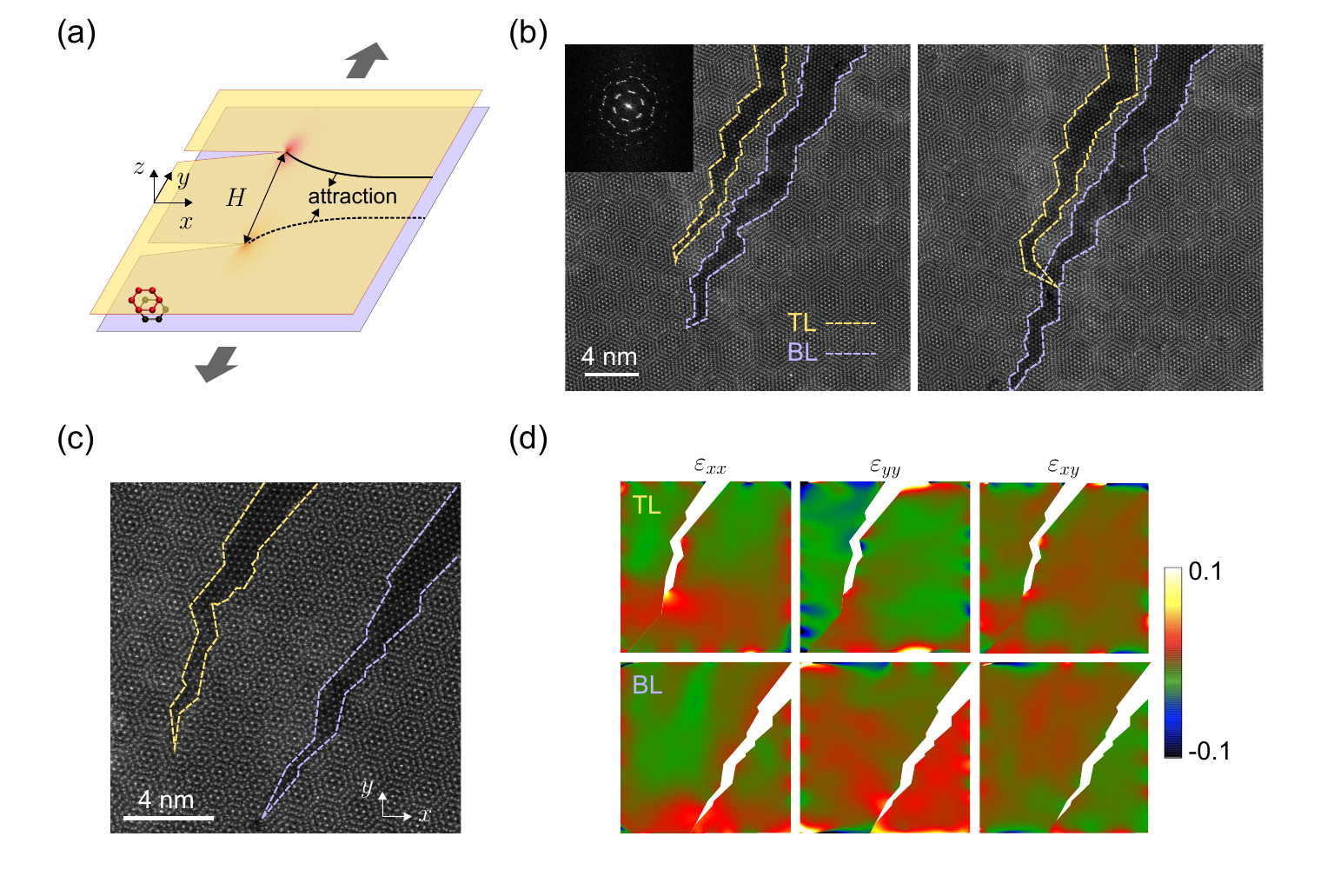}
\caption{\label{fig1} {Attraction between cracks propagating in a bilayer of 2D crystals.}
(a) Schematic illustration of the experimental setup.
(b, c) Cracks in a twisted ${\rm MoS_2}$ bilayer characterized by the atomic-scaled high-angle annual dark field-scanning transmission electron microscope (HAADF-STEM).
TL: top layer.
BL: bottom layer.
Inset: fast Fourier transform (FFT) patterns denoting the twist angle is $8^{\circ}$.
(d) Geometric phase analysis (GPA) shows asymmetric strain distribution around the crack tips in (c).
}
\end{figure}

%\section*{Results}
%\subsection{Experimental observations show the evidence of crack interaction.}
In the experiments, molybdenum sulfide (MoS$_{2}$) bilayers were synthesized through atmospheric pressure chemical vapor deposition (APCVD).
Twisted bilayers were obtained for the weak interaction between the 2D MoS$_{2}$ layers that cannot maintain epitaxial growth.
Precracks were introduced to the bilayers by exposure to ultrahigh intensity \emph{e}-beam ($> 1$ pA/nm$^{2}$), the propagation of which is characterized by atomic-scaled high-angle annual dark field-scanning transmission electron microscopy (HAADF-STEM).
The characterization was operated on the Thermo Fisher Spectra 300 TEM with a field-emission gun at 300 kV~\cite{han2023phase}.
The areas of observation for the crack dynamics are completely outside of the initial area exposed to intensified \emph{e}-beams, therefore experiencing low dose exposure and free from noticeable \textit{e}-beam damage.
The precracks in the two respective layers of a twisted MoS$_{2}$ bilayer are often misaligned due to the anisotropic fracture toughness defined by the lattice symmetry.
However, these cracks tended to approach each other during propagation (Fig.~\ref{fig1}).
%and cross, then depart, allowing the re-bonded of crack edges behind the crack tips, which greatly enhances the fracture toughness [Ref.].
Geometric phase analysis (GPA)~\cite{feng2021strain} on the \textit{in situ} STEM results shows asymmetric strain distributions around the crack tips, which are attributed to the interlayer shear among the stacking area between cracks (Fig.~\ref{fig1}(d)).

%\subsection{Atomistic simulations reveal the effects of interlayer shear.}
To rationalize the interaction between cracks across 2D layers, we carried out molecular dynamics (MD) simulations to probe the dynamics of a pair of cracks using bilayer graphene as a representative model, which can be extended to other 2D crystals with a honeycomb lattice by modifying the force field parameters (see Supplemental Material for details~\cite{supp-info}).
Graphene is chosen for the study instead of MoS$_{2}$ in the experiments because the corresponding force fields of the former are more mature and verified.
%currently there are no chemically accurate force fields developed for the latter.
%A theoretical framework that captures intralayer lattice deformation and interlayer shear or sliding is then presented, offering a predictive model for developing multiple cracks in layered materials.

\begin{figure}[b]
\includegraphics[width=\linewidth]{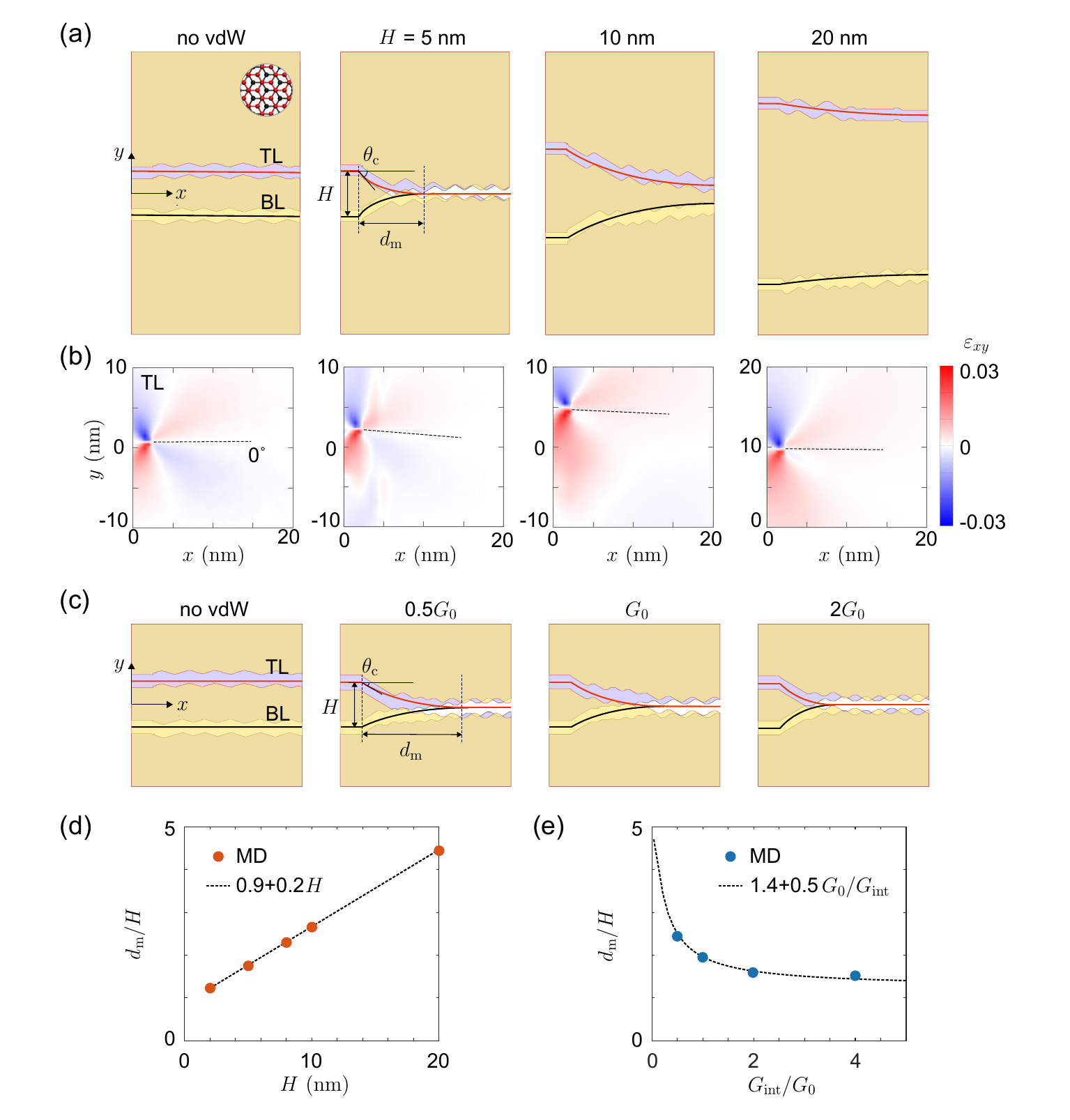}
\caption{\label{fig2} {Fracture patterns of parallel cracks in bilayer graphene simulated by molecular dynamics (MD).}
(a) Crack paths with edge precracks spaced by $H$, obtained from simulations without and with interlayer van der Waals (vdW) interaction.
(b) The atomic shear strain obtained from bond-distortion analysis~\cite{feng2021strain}.
The strain distributions at the crack tips in TL and BL show reflected patterns about the $x$-axis.
The dashed lines denote the axis of symmetry of the atomic shear strain around the crack tip as the crack starts to advance.
(c) Crack paths predicted at different interlayer shear stiffness $G_{\rm int}$ and $H = 5$ nm.
$G_0$ is the interlayer shear stiffness of the graphene bilayer~\cite{liu2012mechanical}.
(d, e) The dependence of $\alpha = d_{\rm m}/H$ on $H$ (d) and $G_{\rm int}/G_{\rm 0}$ (e), where $d_{\rm m}$ is the distance traveled by the cracks before they `merge'.
}
\end{figure}

Our simulation results show that, firstly, interlayer shear partially releases the strain energy and modifies the fracture toughness (Fig. S1~\cite{supp-info}).
The critical mode-I (tensile) stress intensity factor (SIF) or fracture toughness, $K_{\rm I c}$,
%, estimated from the remotely applied stress $\sigma_{y}$ as $K_{\rm I c} \sim \sigma_y \sqrt{\pi a_{0}}$, 
is enhanced by increasing the crack spacing, $H$, and the interlayer shear stiffness $G_{\rm int}$ (Figs. S1(a) and S1(b)~\cite{supp-info}).
%[why $K_{\rm Ic}$ increases with $H$?]
%In a monolayer with cracks, shielding stress between cracks will lower the fracture toughness~\cite{horii1985elastic}.
By comparing the crack configurations and stress concentration around the crack tips with or without the interlayer interaction (Fig. S1(c)~\cite{supp-info}), we find that interlayer shear adds a restoring force, hinders crack opening, and alleviates the in-plane stress concentration at the crack tips.
This cross-layer coupling directly leads to effective toughening.

On the other hand, the fracture patterns show that two parallel cracks in adjacent layers attract each other and eventually `merge' into a single path in the $x$-$y$ plane (Fig.~\ref{fig2}).
The fracture paths display lattice kinks, which are attributed to the lattice discreteness and fracture-resistance anisotropy~\cite{liu2010graphene, kim2012ripping, shi2023nonequilibrium}.
By ignoring the kinking features, the strength of attraction is assessed through the crack deflection angle ($\theta_{\rm c}$) from the distance traveled by the cracks before they merge ($d_{\rm m}$) (Fig.~\ref{fig2}(a)).
The value of $\theta_{\rm c}$ at the coarse scale ($\sim d_{\rm m})$ is related to the geometric factor $\alpha = d_{\rm m}/H$ as ${\rm tan}(90^{\circ}-\theta_{\rm c}) \sim \alpha$.
The simulation results show that $\alpha$ increases linearly with $H$ (Fig.~\ref{fig2}(d)), suggesting that $\theta_{\rm c}$ is inversely related to $H$.
The attraction is thus weakened as the overlap area between the cleaved edges of the two cracks increases.
The attraction between cracks is absent by turning off the interlayer interaction.

The strain distribution in the bilayer is analyzed by bond-distortion analysis~\cite{feng2021strain}.
As the crack starts to propagate, the symmetry to the crack direction (the $x$ direction) is broken and gradually recovered for configurations with large values of $H$ (Fig.~\ref{fig2}(b)).
A mode-II SIF (shear, $K_{\rm II}$) emerges at the crack tips even under mode-I loading as the result of the interlayer interaction.
The control of crack paths through $K_{\rm I}$ and $K_{\rm II}$ can be rationalized, for example, through the criterion of maximum energy release rate (MERR) criterion~\cite{feng2022experimentally}.
The crack deflection at the coarse scale depends on the competition between the driving force 
%(measured by the ratio or mode-mixing factor, $\beta = K_{\rm II} / K_{\rm I}$)
and the material anisotropy in fracture resistance defined by the crystal symmetries as evidenced by the kinking features.
Consistently, experimental observation identifies asymmetric stress at the crack tips of parallel cracks, modifying their advancing directions (Fig.~\ref{fig1}(d)).
In addition, the effect of interaction strength is studied by rescaling all the energy-related coefficients ($\epsilon, C, C_6$) of ILP potential~\cite{ouyang2018nanoserpents}.
The results show that the strength of attraction between cracks increases with $G_{\rm int}$ (modified from the original value of  $G_0$), and the ratio $\alpha = d_{\rm m}/ H$ decreases as a result (Figs.~\ref{fig2}(c) and \ref{fig2}(e)).
These MD simulation results uncover the underlying mechanisms with an atomic-level picture.

\begin{figure}[t]
\includegraphics[width=\linewidth]{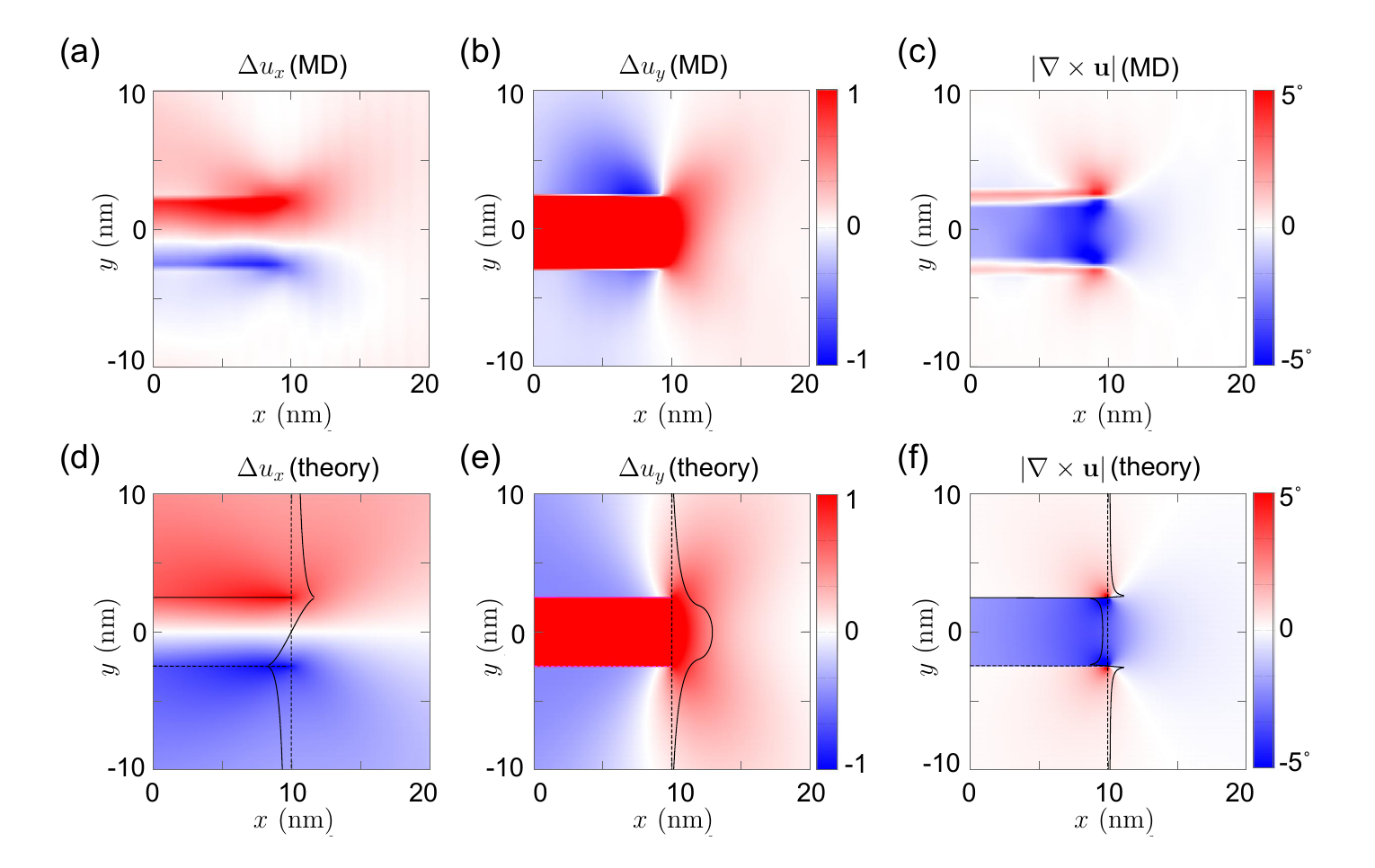}
\caption{\label{fig3} {Interlayer slip distribution.}
(a-f) Interlayer displacements, $\Delta u_{x}$ and $\Delta u_{y}$, and the curl, $\lvert\nabla \times \mathbf{u}\rvert$, predicted by MD simulations (a-c), and by continuum theory analysis (d-f) for precracks with $H = 5$ nm, respectively. 
The lines in panel B show the distribution of $\Delta u_{x}$, $\Delta u_{y}$, and $\lvert\nabla \times \mathbf{u}\rvert$ along the $y$-axis at the crack tip.
}
\end{figure}

%\subsection{Theoretical models predict the experimental and simulation results.}
The stress focusing at the crack tip yields highly non-uniform displacement fields.
For bilayers containing one crack each, the difference between in-plane displacement fields in the two layers results in interlayer shear.
%From our MD simulations at specific configurations and before the propagation of cracks
The interlayer slip $\Delta {\bf u}$ of the graphene bilayer can be characterized in the unit of the C-C bond length $b = 0.142$ nm.
Figs.~\ref{fig3}(a)-\ref{fig3}(c) shows localized slips between two parallel cracks spaced by $H = 5 $ nm.
The slip amplitudes $\Delta u_x$ and $\Delta u_y$ are concentrated around the crack tips, which are anti-symmetric and symmetric around the axis $x=0$, respectively. 
Linear elastic fracture mechanics (LEFM) analysis (Supplemental Material~\cite{supp-info}) shows that the spatial distribution of $\Delta {\bf u}$ agrees well with the MD simulation results (Figs.~\ref{fig3}(d)-\ref{fig3}(f)).
The amplitude of slip is slightly lower in the simulations due to the self-adjustment of lattice registries~\cite{zhang2022domino, yoo2019atomic}, which is not included in LEFM.
%as discovered in the $\sim 1^{\circ}$ twisted graphene bilayer.

\begin{figure}[b]
\includegraphics[width=\linewidth]{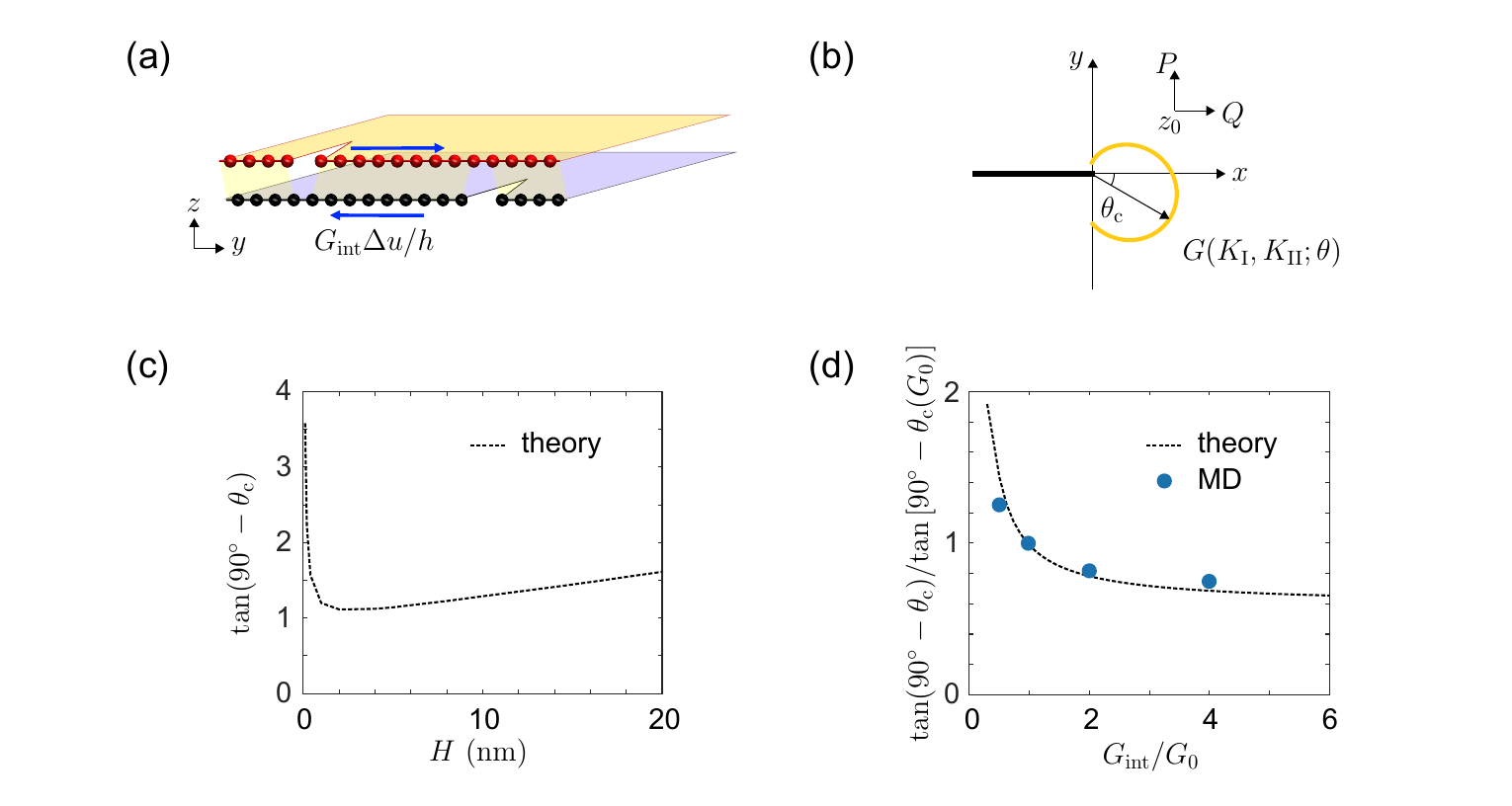}
\caption{\label{fig4} {Theoretical prediction of crack interaction.}
(a) The shear-lag model captures the load transfer between neighboring layers.
(b) Calculations of the stress intensity factors (SIFs, $K_{\rm I}, K_{\rm II}$) under a concentrated force ($P, Q$) at $z_0 = x_0 + i y_0$ measured from the crack tip.
The incomplete curve around the crack tip shows the distribution of $G(\theta)$ under $K_{\rm I}, K_{\rm II} (\ne 0)$.
The direction of crack propagation is determined by $\theta_{\rm c}$ = ${\rm max} G(\theta)$ according to the criterion of maximum energy release rate (MERR).
(c) The strength of attraction, ${\rm tan} (90^{\circ} - \theta_{\rm c})$, plotted as a function of $H$.
(d) The normalized strength of attraction, ${\rm tan} (90^{\circ} - \theta_{\rm c}) / {\rm tan} \left[90^{\circ} - \theta_{\rm c}(G_0)\right]$, plotted as a function of $G_{\rm int} / G_0$ and compared to the MD simulation results.
}
\end{figure}

The discrete nature of 2D crystals results in the formation of Moir\'e patterns in the bilayers, the distortion of which indicates the lattice strain.
Both experimental and simulation results show that crack opening results in relative twisting of the bilayers.
%The amount of twist can be estimated from the periods of the Moire patterns, $L_{\rm m}$, the distortion of which indicates the lattice strain.
%In \texttt{AB}-stacked bilayers, the Moir\'e patterns form only in the region between cracks, and the crack tips are always in \texttt{AA} stacking during crack advancing, where the twist is the most significant (Fig.~\ref{fig1}c and Fig.~\ref{fig3}).
%In twisted bilayers in the absence of strain, the periods of the Moir\'e patterns inversely correlate with the twist angle ($\theta$) as $L_{\rm m} = a/2 {\rm sin}(\theta/2)$, where $a$ is the lattice constant of graphene.
%The period of Moir\'e patterns in the region between the two cracks is smaller than that of those in other regions (Fig.~\ref{fig1}c), which measures the local strain around the crack tips~\cite{kazmierczak2021strain}.
By ignoring the lattice registry at the interface, we can further assume that the interlayer interaction follows a linear constitutive shear model.
This approximation remains valid before interlayer sliding is activated and $|\Delta {\bf u}| < b$.
The interlayer shear stiffness $G_{\rm int}$ can then be treated as a constant, and the interlayer shear stresses are $\tau_{xz} = G_{\rm int} \Delta u_{x} /h, \tau_{yz} = G_{\rm int} \Delta u_{y} /h$ (Fig.~\ref{fig4}(a)), where $h = 0.335$ nm is the interlayer distance.
A shear-lag analysis (Supplemental Material~\cite{supp-info}) is then used to calculate the stress intensity factors (SIFs, $K_{\rm I}$ and $K_{\rm II}$).

The direction of crack deflection $\theta_{\rm c}$ at the coarse scale can then be determined from MERR (Supplemental Material~\cite{supp-info}).
By using the interlayer shear modulus of graphene ($G_{\rm 0} = 3.48$ GPa)~\cite{liu2012mechanical}, we find that the value of ${\rm tan} (90^{\circ} - \theta_{\rm c})$ decreases with $H$ for $H < 2$ nm, and then increases linearly with $H$ (Fig.~\ref{fig4}(c)).
The analytical prediction agrees well with the MD simulation results summarized in Fig.~\ref{fig2}(d). 
Specifically, for cracks that align closely with $H < 2$ nm, the overlap area between the crack tips is small even if the amplitudes of relative in-plane displacement and interlayer shear are large.
As a result, the attraction is weak, and no deflection is observed in the MD simulations.
For $H > 2$ nm, both $K_{\rm I}^{\rm int}~(<0)$ and $K_{\rm II}^{\rm int}~(>0)$ decrease with $H$, so does the ratio between overall SIFs, $\beta$, which results in crack deflection with an angle diminishing as $H$ increases.
%\zp{The maximum crack deflection angle is $\theta \sim 45^{\circ}$.}
In addition, the normalized value of ${\rm tan} (90^{\circ} - \theta_{\rm c})$ decreases with $G_{\rm int}$ in a logarithmic form, in consistency with the MD simulation results (Fig.~\ref{fig4}(d)).

\begin{figure}[t]
\includegraphics[width=\linewidth]{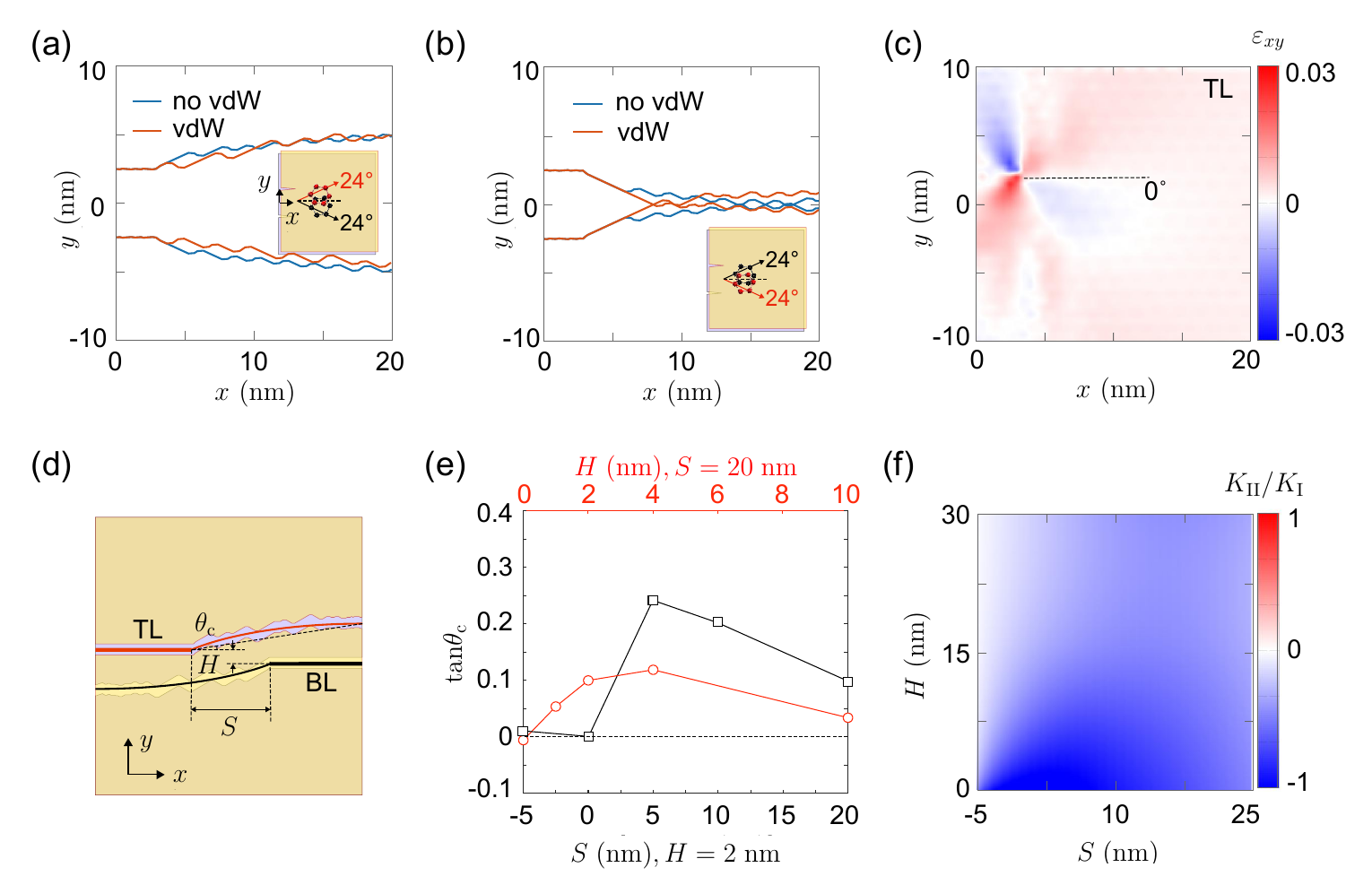}
\caption{\label{fig5} {Interaction between twisted graphene bilayer cracks and \emph{En-Passant} cracks in a bilayer.}
(a, b) Fracture patterns of parallel cracks obtained from MD simulations with/without interlayer interaction ($H = 50$~nm).
The insets show the geometries, where the angles between the zigzag directions and the $x$-axis in TL and BL are $24^{\circ}$ and $-24^{\circ}$, respectively.
(c) Shear strain at the crack tip in TL of the twisted bilayer (corresponding to the simulations for panel A, with vdW interaction), which is symmetric about the crack direction.
(d) Fracture patterns of anti-parallel cracks advancing toward each other.
Here $H$ and $S$ are the initial distances between the crack tips in the $x$- and $y$-directions, respectively.
(e) The value of ${\rm tan} \theta_{\rm c}$ calculated from the fracture patterns as functions of $H$ and $S$, showing repulsion in \emph{En-Passant} cracks (${\rm tan} \theta_{\rm c} > 0$).
(f) Theoretical predictions of the mode-mixing factor $\beta = K_{\rm II} / K_{\rm I}$ at the crack tip.
The negative values of $\beta$ suggest anti-clockwise deflection, in consistency with the MD simulation results.
}
\end{figure}

%\section*{Discussion}
The interlayer load transfer by shear between 2D crystals depends on the twist angle.
%The shear between graphene bilayers becomes much reduced while deviating from the \texttt{AB} stacking \zp{[Ref.]}.
MD simulations with a twist angle of $\theta = 12^{\circ}$ show no obvious cracks deflection from their intrinsic directions of propagation defined by the loading condition and the crystal orientation-dependent fracture toughness~\cite{shi2023nonequilibrium} (Figs.~\ref{fig5}(a) and \ref{fig5}(b)).
Compared to the bilayers in \texttt{AB} stacking (Fig.~\ref{fig2}(b)), shear strain at the crack tip in the twisted bilayer is symmetric about the crack direction (Fig.~\ref{fig5}(c)), suggesting that interlayer sliding has negligible contribution to the in-plane lattice distortion.
The effect of twist angles on the fracture of bilayers can be further discussed by tuning the shear stiffness $G_{\rm int}$ in the shear lag model.
For graphene, the shear stiffness of twisted bilayers is significantly lower than that in AB-bilayers~\cite{hod2018structural}, and the mode-mixing factor $\beta = K_{\rm II} / K_{\rm I}$ is negligible.
The loading conditions and intrinsic anisotropic fracture toughness of 2D crystals determine the crack direction.
For MoS$_2$ with a higher interlayer shear modulus of $G_{\rm int} = 16.4$~GPa~\cite{zhao2013interlayer} and lower in-plane modulus of $E = 270 \pm 100$~GPa~\cite{bertolazzi2011stretching}, the shear responses were reported to be independent of the twist angle, possibly due to the low in-plane stiffness that allows more significant in-plane lattice relaxation favoring local registry match ~\cite{sun2022determining}.

Our discussion also extends to crack configurations with different crack lengths and directions of propagation in 2D bilayers.
For parallel cracks with different lengths in bilayer \texttt{AB}-stacking graphene, attraction is prevalent in parallel cracks (with the same direction of propagation, Fig. S3~\cite{supp-info}).
The shorter crack always propagates ahead of the longer one, and the two cracks chase each other and take the lead alternately since the critical SIF is $K_{\rm Ic} \sim \sigma_y \sqrt{\pi a_{0}}$, where $\sigma_y$ is lower for short cracks.
For anti-parallel or `\emph{En-Passant}' cracks in the bilayers, 
%where the precracks are located at opposite edges and tend to extend towards each other without any interaction. But with interlayer interaction included, 
the cracks avoid encountering each other during propagating, demonstrating repulsion between the cracks (Fig. S4~\cite{supp-info}).
The repulsive strength relies on the distances between the two crack tips (Fig.~\ref{fig5}(d)), where $H$ and $S$ are the initial distances between the crack tips in the $x$- and $y$-directions, respectively.
The deflection angle $\theta_{\rm c}$ is measured from MD simulations, showing the repulsion becomes stronger as the cracks approach each other firstly and then vanishes as the values of $H$ or $S$ become close to $0$ (Fig.~\ref{fig5}(e)).
Moreover,  the interlayer slips calculated from MD simulations and continuum theory analysis show opposite directions of $\Delta {\bf u}$ around the crack tips, in contrast to the parallel cracks (Figs.~\ref{fig3} and S5~\cite{supp-info}).
By integrating the SIFs over the crack-tip domain,
%using $\Delta {\bf u}$ calculated from Eq.~\ref{eq-3}, 
we find that $\beta < 0$ at the crack tip is preserved.
This result validates the repulsion between anti-parallel cracks in the bilayer and its enhancement as $H$ and $S$ decrease (Fig.~\ref{fig5}(f)).

For comparison, the interaction between co-planar cracks is also studied using finite element analysis (FEA, Fig. S6~\cite{supp-info}), which can be inferred from the mode-mixing factor $\beta = K_{\rm II} / K_{\rm I}$.
Positive or negative $K_{\rm II} / K_{\rm I}$  values result in clockwise or anti-clockwise crack deflection, and thus attraction or repulsion, respectively.
For parallel cracks with identical lengths, the cracks tend to repel each other (Figs. S6(a) and S6(b)~\cite{supp-info}), while for those with length contrast, the interaction depends on their crack-tip distances (Figs. S6(c) and S6(d)~\cite{supp-info}). 
The longer crack advances by experiencing repulsion and slight attraction by the neighboring shorter one as the longer crack extends, while the shorter one tends not to advance due to stress shielding.
For \emph{En-Passant} edge cracks, the calculated SIFs show that, when the cracks extend towards each other, the shear component $K_{\rm II}$ transits from negative to positive after the two crack tips encounter (Figs. S6(e) and S6(f)~\cite{supp-info}), explaining the attraction-after-repulsion behaviors.
Co-planar cracks interact by experiencing crack-tip stress shielding or amplification that depends on the crack configurations.
The cracks curve as a result to minimize the shear stress and maximize the energy release rate~\cite{dalbe2015repulsion}.
However, for cracks in neighboring layers of a bilayer, the direction of interlayer slip depends on the crack directions only.
The in-plane shear around the crack tips is always in the same direction, and as a result, only attraction (repulsion) is observed for parallel (anti-parallel) cracks.

%\section*{Conclusion}
In summary, we demonstrate the control of crack paths through interaction across neighbouring 2D layers by atomic-scale \emph{in situ} electron microscopy and molecular dynamics simulations.
The interaction that defines the crack paths is attributed to interlayer shear resulting from non-uniform deformation around the crack tips.
The phenomenon is a direct consequence of the layered anisotropy of 2D crystals and is well predicted by continuum mechanics analysis by incorporating the linear elasticity of fracture mechanics and the shear-lag model.
%These features make 2D materials a platform to explore the fracture mechanics of layered materials~\cite{freund2004thin}.
The findings can guide the material design for engineered fracture patterns and modified crack resistance.
For example, by harnessing the attraction between parallel cracks in neighboring layers, the `merging' of precracks can improve fracture resistance, provided that the interlayer interaction can be strengthened.

%at for cracks propagating within adjacent layers in 2D crystals, the interlayer shear mediates a coupling between the stress field in the layers and results in an effective attraction or repulsion between the cracks.
%Attraction between parallel cracks in the same direction, and repulsion between \emph{En-Passant} cracks in the opposite direction.
%The interaction strength increases with the interlayer shear modulus and decreases with the spacing between cracks.
%The conclusion can be extended to bilayers with a weaker interlayer registry.
%Our simulation results show that for layers that are not in AB stacking, the load transfer is insufficient, and thus the interaction between cracks in neighboring layers is not obvious.
%The problem solved in this work demonstrates the integration between fracture and interlayer load transfer.
%In laminates with strong adhesion, large stress concentration at the crack tip is fully transferred to the other layer, and the crack tends to go cross the interface.
%While with weak adhesion, interlayer shear can easily reach its maximum before crack propagation, and then interface delamination occurs first.
%However, the vdW stiffness is degrees lower than the intralayer elastic stiffness, limits the interlayer loading transfer in 2D crystals, and delamination will never happen due to the restorability of the vdW interaction.

\begin{acknowledgments}
We wish to acknowledge the financial support by the National Natural Science Foundation of China through grants 11825203, 52090032, and 52173230, the National Key Basic Research Program of China grant No. 2022YFA1205400, the China Postdoctoral Science Foundation through the grant 2023M731967, the Hong Kong Research Grant Council General Research Fund through grants 15302522, 15301623.
The computation was performed on the Explorer 1000 cluster system of the Tsinghua National Laboratory for Information Science and Technology.
S.F. and X.Z. contributed equally to this work.
\end{acknowledgments}

%$^{\ast}$ jiongzhao@polyu.edu.hk\\
%$^{\dagger}$ xuzp@tsinghua.edu.cn\\

%\nocite{*}
\bibliography{main_text}% Produces the bibliography via BibTeX.

\end{document}